\documentclass[fleqn,usenatbib]{mnras}

\usepackage{newtxtext,newtxmath}

\usepackage[T1]{fontenc}
\usepackage{ae,aecompl}

\usepackage{graphicx}	
\usepackage{bm}
\usepackage{xspace}

\newcommand{\cf}{cf.,~}
\newcommand{\ie}{i.e.,~}
\newcommand{\eg}{e.g.,~}

\usepackage{todonotes}
\newcommand{\bhac}{\texttt{BHAC}\xspace}

\title[3D MHD jet break out]{3D magnetised jet break-out from
  neutron-star binary merger ejecta: afterglow emission from the jet and
  the ejecta}

\author[A. Nathanail et al.]{Antonios Nathanail$^{1}$, Ramandeep
  Gill$^{2,3}$, Oliver Porth$^{4}$, Christian M. Fromm$^{1,5}$, and \newauthor
  Luciano Rezzolla$^{1,6,7}$ \\
$^{1}$Institut f\"ur Theoretische Physik, Goethe Universit\"at Frankfurt,
Max-von-Laue-Str.1, 60438 Frankfurt am Main, Germany \\
$^{2}$Department of Physics, The George Washington University, 
  Washington, DC 20052, USA \\
$^{3}$Department of Natural Sciences, The Open University of 
  Israel, 1 University Road, PO Box 808, Raanana 4353701, Israel \\
  $^{4}$Astronomical Institute Anton Pannekoek, Universeit van Amsterdam, 
  Science Park 904, 1098 XH, Amsterdam, The Netherlands \\
$^{5}$Max-Planck-Institut f\"ur Radioastronomie, Auf dem H\"ugel 69,
D-53121 Bonn, Germany \\
$^{6}$Helmholtz Research Academy Hesse for FAIR, Max-von-Laue-Str. 12,
60438 Frankfurt am Main, Germany\\ 
$^{7}$School of Mathematics, Trinity College, Dublin 2, Ireland
}

\usepackage{hyperref}
\hypersetup{
  colorlinks=true,        
  linkcolor=blue,         
  citecolor=cyan,      
  filecolor=magenta,      
  urlcolor=magenta            
}

\begin{document}
\label{firstpage}
\maketitle

\begin{abstract}
We perform three-dimensional (3D) general-relativistic
magnetohydrodynamic simulations to model the jet break-out from the
ejecta expected to be produced in a binary neutron-star merger. The
structure of the relativistic outflow from the 3D simulation confirms our
previous results from 2D simulations, namely, that a relativistic
magnetized outflow breaking out from the merger ejecta exhibits a hollow
core of $\theta_{\rm core}\approx4^{\circ}$, an opening angle of
$\theta_{\rm jet}\gtrsim10^{\circ}$, and is accompanied by a wind of
ejected matter that will contribute to the kilonova emission. We also
compute the non-thermal afterglow emission of the relativistic outflow
and fit it to the panchromatic afterglow from GRB170817A, together with
the superluminal motion reported from VLBI observations. In this way, we
deduce an observer angle of $\theta_{\rm obs}= 35.7^{\circ
  \,\,+1.8}_{\phantom{\circ \,\,}-2.2}$. We further compute the afterglow
emission from the ejected matter and constrain the parameter space for a
scenario in which the matter responsible for the thermal kilonova
emission will also lead to a non-thermal emission yet to be observed.
\end{abstract}




\section{Introduction}
\label{sec:intro}

Binary neutron-star (BNS) mergers offer a rich variety of observables,
from the gravitational waves (GWs) and kilonova emission powered by the
neutron rich ejected matter to the prompt emission from an
ultra-relativistic jet and its late-time afterglow signal. A short
gamma-ray burst (GRB) is expected after the jet breaks out from the BNS
merger ejecta. The first ever detection of GWs from a BNS merger,
GW170817 \citep{Abbott2017}, was accompanied by a number of
electromagnetic signals \citep{Abbott2017b}: a coincident detection of a
short GRB, GRB170817A \citep{Savchenko2017,Goldstein2017}, a
quasi-thermal kilonova detection from the first day till several days
later \citep{Arcavi2017, Nicholl2017, Chornock2017, Cowperthwaite2017,
  Villar2017}, and a non-thermal afterglow emission that appeared in the
X-rays nine days after the merger \citep{Troja2017}, and in the optical,
and radio bands more than fifteen days post-merger
\citep{Hallinan2017}. The latest afterglow observations were in the
X-rays at $743$ and $940$ days after the merger
\citep{Hajela2019,Makhathini2020}.

To understand the outflow energetics and its structure, the afterglow was
extensively studied from the continuous brightening of the flux, with its
peculiar shallow rise to the peak at $t_{\rm pk}\simeq150\,{\rm d}$
post-merger \citep{Lyman2018,Margutti2018,Mooley2018}, to well into the
late-time decay phase after the lightcurve peak
\citep{Alexander2018,Troja2019}. Several works developed simplified
semi-analytical models and employed two-dimensional (2D) nonlinear
numerical simulations to understand the shallow rise of the afterglow
lightcurve before its peak. Two distinct types of outflow structures were
tested against the observations: one featured a ``structured jet'', with
a polar structure and a narrow relativistic core surrounded by low-energy
wings \citep[e.g., ][]{Troja2017, Troja2018, Gill2018, DAvanzo2018,
  Margutti2018, Lazzati2017c, Lamb2018, Pozanenko2018, Beniamini2019},
and the second featured a ``cocoon'', namely, a wide-angle outflow
expanding quasi-spherically, with a radial velocity stratification
\citep[e.g.,][]
      {Kasliwal2017,Gill2018,Gottlieb2018,Salafia2018a,Mooley2018,Fraija2018}.
      The radio observation of an apparent superluminal motion of the
      flux centroid \citep{Mooley2018b}, together with a compact size
      (\ie $\lesssim2\,$mas) of the outflow on the plane of the sky
      \citep{Ghirlanda2019}, favoured the structured jet model, that was
      needed to explain the afterglow lightcurve and image properties
      near and after the peak \citep{Lamb2018b}.

The details of the afterglow lightcurve \citep[e.g.,][]{Beniamini2020b}
and image properties depend mainly on the jet structure, the ambient
rest-mass density and our line-of-sight.  The structure of the jet, on
the other hand, is determined already early on at the site where the jet
is launched, which is expected to happen in the vicinity of a rotating
black hole, but also through its interaction with the circumburst medium
through which the jet is drilling a hole. Such a medium is composed of
the matter ejected during the merger -- either dynamically or secularly
\cite[see, \eg][for some reviews]{Baiotti2016,Paschalidis2016} -- but
also of the accretion torus that is formed when the merger remnant
collapses to a black hole. Both of these components affects the jet's
structure and propagation.

Hydrodynamic jets have been studied analytically in a large number of
recent works \citep{Bromberg2011, Nakar2017, Lazzati2017c, Salafia2018a}
and numerically \citep{Nagakura2014, Murguia-Berthier2014,
  Murguia-Berthier2016, Lamb2017a, Duffell2015, Duffell2018,
  Lazzati2017b, Xie2018, Gottlieb2018b, Matsumoto2019, Hamidani2020,
  Hamidani2020b, Gottlieb2020}. On the other hand, studies of (MHD) jets
in the context of short GRBs and produced after a BNS merger are
considerably fewer \citep{Kathirgamaraju2018, Bromberg2018, Geng2019,
  Fernandez2018, Kathirgamaraju2019, Sapountzis2019, Nathanail2020b,
  Gottlieb2020b}. Yet, there are substantial differences between the
properties of a hydrodynamic jet and of an MHD jet breaking out from the
merger ejecta.

By carrying out a series of 2D of general-relativistic MHD (GRMHD)
simulations that considered variations in the matter distribution around
a black hole and its spin as well as the magnetic field geometry and
energy, \citealt{Nathanail2020b} suggested that MHD jets have the
following properties: \textit{(i)} they have an intrinsic polar structure
for the energy and velocity with a ``hollow core'' of $\theta_{\rm
  core}\approx4^{\circ}-5^{\circ}$ (measured from the jet symmetry axis)
and an opening angle of $\theta_{\rm jet}\gtrsim10^{\circ}$;
\textit{(ii)} they are followed by a disc wind with significant amounts
of matter, which contributes to the overall ejected mass that follows a
BNS merger; \textit{(iii)} they naturally develop an angular (and radial)
energy stratification by the interaction of the surrounding matter.

The afterglow of GRB170817A was analysed mainly by using numerical and
semi-analytical models of hydrodynamic jets that best fit the afterglow
data, and correspond to structured jets with a relativistic core of
angular size $\theta_{\rm core}\sim 3^{\circ}-5^{\circ}$
\citep{Resmi2018,Mooley2018b,Ghirlanda2019,Troja2019,Lamb2019,
  Beniamini2018,Gill2019b,Nakar2020}. However, in the case of an MHD jet,
the opening angle is certainly very different \citep{Kathirgamaraju2019,
  Nathanail2020b}\footnote{Note that weakly magnetized jets do not show
the characteristic structure of MHD jets \citep{Gottlieb2020b}.}, which
would affect also the resulting viewing angle of the afterglow emission.

In this study we report three-dimensional (3D) GRMHD simulations to
explore the structure of an MHD jet as it breaks out the merger
ejecta. The jet is launched self-consistently after the massive remnant,
produced from the BNS merger, collapses to a black hole. Overall, our
simulations confirm our previous results on the intrinsic properties of
MHD jets that were found from 2D GRMHD simulations
\citep{Nathanail2020b}. We further explore the MHD wind that accompanies
the jet, another source of ejected mass, by calculating its non-thermal
afterglow emission and investigating possibilities for its observational
signature.

This paper is organised as follows: in Section \ref{sec:comp} we present
the numerical setup and the main results for the structure of the 3D MHD
jet. In Section \ref{sec:after}, we discuss the method used to compute
the afterglow emission and present the resulting lightcurves. The
details of the ejected mass produced by the MHD wind from the accretion
disc are presented in Section \ref{sec:eject}, together with the
non-thermal observable properties of this component. Finally, we
summarise and present a discussion about the results in Section
\ref{sec:con}.

\section{MHD jets in 3D}
\label{sec:comp}

For our 2D and 3D simulations, we employ \bhac to solve the
general-relativistic MHD equations in a Kerr background spacetime
\citep{Porth2017,Olivares2019}, in the ideal-MHD limit
\citep{Harutyunyan2018}. To model the torus around the compact remnant
together with the ejected matter that followed a BNS merger, we follow
the setup introduced in \citet{Nathanail2018c}. We consider a non
self-gravitating torus with initial size $r_{\rm in}=6\, M=23.8\,{\rm
  km}$ and $r_{\rm max} = 14.3\, M =56.7\,{\rm km}$
\citep{Fishbone76,Abramowicz78} around a remnant black hole of mass $M =
2.7\, M_{\odot}$, with a dimensionless spin of $a:=J/M^2=0.9375$.  In our
previous study we showed that the spin parameter does not have a strong
impact on the jet structure, more specifically lower spins generate jets
that break out from the merger ejecta \citep{Nathanail2020b}. We recall
that numerical-relativity simulations reveal that the BH produced from a
BNS merger is expected to have a dimensionless spin $a \sim 0.7-0.9$
\citep{Kastaun2013, Bernuzzi2013, Baiotti2016}.
We should remark that adopting a low-spin prior for GW170817
  implies that the most realistic value for the dimensionless spin is $a
  \simeq 0.8$ \citep{Kastaun2013, Bernuzzi2013}. In turn, assuming that
  the power in the jet scales quadratically with $a$ -- as in the
  Blandford-Znajek mechanism \citep{Blandford1977} -- reveals that the
  jet power in our simulations is overestimated by $\sim 23\%$, which is
  well within the systematic error budget of our modelling. More
  importantly, the structure of the jet is not significantly affected by
  the exact value of $a$ as long as the jet energy is considerably larger
  than that in the ejecta. The jet structure is indeed affected in cases
  where the jet barely makes out of the ejecta \citep[e.g.,][for
    hydrodynamical simulations of jet breakout for different jet powers
    and durations]{Urrutia2020}.
  
 The initial matter distribution has a radial extent of $1200\,{\rm km}$,
 in order to account for the expanded torus, and also for the dynamically
 ejected matter during merger that has reached such a distance, since the
 launch of a relativistic jet from GW170817/GRB170817A is expected to be
 significantly later from merger after the remnant collapsed to a black
 hole, \ie $\approx 1\,{\rm s}$
 \citep{Margalit2017,Shibata2017c,Rezzolla2017,Granot2017,
   Nathanail2018,Gill2019,Beniamini2020,Murguia-Berthier2020}. The
 numerical domain has a radius of $10^4\,{\rm km}$ and initially all
 matter is bound with having zero radial velocity. Here we report a 3D
 simulation with effective resolution of $192^3$ realised by three
 refinement levels and as a comparison, we report another two simulations
 performed in axisymmetry (2D) with a resolution of $1024\times 512$
 using three refinement levels.

An important point of the initial post-merger configuration is the
non-empty funnel in the polar region, which is expected as a result of
the dynamically ejected matter during the merger
\citep{Sekiguchi2016,Foucart2016a,Radice2016,
  Bovard2017,Dietrich2017,Fujibayashi2017b,Most2019b}. To model this, we
fill the polar region with matter that has a $2.5$ orders of magnitude
less rest-mass density than the maximum rest-mass density of the torus
and its radial profile scales like $r^{-1.5}$, $\rho_{\rm polar}(r) =
10^{-2.5}\times\rho_{\rm max} (r/r_{\rm max})^{-1.5}$. Initially,
  this matter distribution has a zero radial velocity, since the relative
  average velocity of these ejecta is
  $\langle\beta\rangle:=\langle\,v\,\rangle/c\sim0.1-0.2$~
  \citep{Rezzolla:2010, Siegel2014, Sekiguchi2016, Foucart2016a,
    Radice2016, Bovard2017, Dietrich2017, Fujibayashi2017b, Most2019b}
  and it is considerably slower than the generated relativistic
  outflow. Given that our focus here is the angular structure of the
  relativistic jet as it traverses the dynamical ejecta -- which is most
  sensitive to the density profile of the ejecta -- we believe that this
  is reasonable approximation. On the other hand, a sub-relativistic
  velocity of the ejecta is important for calculating, \eg the jet
  breakout timescale, as indicated by some recent analytic work
  \citep{Hamidani2020}.

During merger, the initial magnetic field of the two neutron stars is
amplified in the first milliseconds via instabilities such as
the Kelvin-Helmholtz or the magnetorotational (MRI), yielding a magnetic
energy for the remnant $>10^{50}\,{\rm erg}$, with a dominant toroidal
component with a factor of $\approx 3$ larger than the poloidal one
\citep{Giacomazzo:2014b,Palenzuela2015,Kiuchi2015a,Kiuchi2017}. Thus the
initial magnetic energy is $\leq10^{50}\,{\rm erg}$, which imposes a
magnetisation in the torus. The magnetic field generated by these
  instabilities is mostly poloidal and on small scales
  \citep{Siegel2013}, although also a global toroidal component with a
  similar magnetic energy is achieved $\approx 20 {\rm ms}$ after the
  merger \citep{Ciolfi2017, Most2019b}.

The mass in the torus is $M_{\mathrm{tor}}\approx 0.1 \, M_{\odot}$
whereas the mass surrounding the torus, which initially is bound for all
simulations, is $M_{\mathrm{surrounding}}=0.02 M_{\odot}$. We initialise
a poloidal nested-loop magnetic field, together with a toroidal component
that traces the fluid pressure, with the desired ratio of $\approx 3$ in
the corresponding magnetic energies. The details of the models presented
in this study are shown in Table \ref{tab:initial}.

\begin{table*}
  \caption{Properties of the three MHD jets
    considered: magnetic energy ($E_B$), average magnetisation in the
    torus ($\sigma:=B^2/4\pi \rho$), maximum rest-mass density of the
    torus ($\rho_{\mathrm{max}}$), initial total mass
    ($M_{\mathrm{tot}}=M_{\mathrm{tor}}+ M_{\mathrm{surrounding}}$, where
    $M_{\mathrm{tor}}$ is the initial torus mass and
    $M_{\mathrm{surrounding}}=0.02 M_{\odot}$ is the mass surrounding the
    torus which initially is bound and in all simulations), the unbound
    ejected mass at the end of the simulation ($M_{\mathrm{ej}}$) and
    their ratio.}  \centering
  \begin{tabular}{|l||c|c|c|c|c|c|c|c|c|c|c|c|c|c|}
    \hline
    \hline
    \!\!\!  model&
    $ E_{B}$& $\overline{\sigma}$&$\rho_{\mathrm{max}}$&  $M_{\mathrm{tot}}$ & $M_{\mathrm{ej}}$ &$\frac{M_{\mathrm{ej}}}{M_{\mathrm{tot}}}$ \\
    \!\!\! & $[\mathrm{erg}]$ & 
     &\!\!\! $[\mathrm{g/cm^3}]$ & $[M_{\odot}]$& $[M_{\odot}]$& $\%$\\
     &$10^{49}$&$10^{-6}$&$10^{10}$&&
     &&\\
  \hline
\! \texttt{MHD-3D-HB} &$10.0$ &$3.9$   & $45$ & $0.109$& $0.027$& $26.1$ \\
\! \texttt{MHD-2D-HB} &$9.0$  &$0.065$ & $2.0$ & $0.144$& $0.038$&$27.1$\\
\! \texttt{MHD-2D-MB} &$1.2$  &$0.036$ & $1.5$ & $0.108$& $0.014$&$14.1$  \\
\hline
\hline
  \end{tabular}  
  \label{tab:initial}
\end{table*}

In our previous study we explored a wide range of parameters for the
initial configuration by varying the initial torus size and shape, the
magnetic field strength and ratio (poloidal to toroidal component), and
the BH spin \citep{Nathanail2020b}. The main results of that parameter
study was that the MHD jets, breaking out from the BNS merger ejecta,
have several intrinsic properties that depend insignificantly on the
initial configuration: \textit{(i)} they have an intrinsic polar
structure for the energy and velocity with a ``hollow core'' of
$\theta_{\rm core}\approx4^{\circ}-5^{\circ}$ (measured from the jet
symmetry axis) and an opening angle of $\theta_{\rm
  jet}\gtrsim10^{\circ}$; \textit{(ii)} they are followed by a disc wind
with significant amounts of matter, which contributes to the overall
ejected mass that follows a BNS merger; \textit{(iii)} they naturally
develop an angular (and radial) energy stratification. All our
simulations were performed in 2D, thus there was a need to continue this
analysis in three dimensions to explore the impact of known instabilities
that occur in 3D. One example is the ``plug'' of heavy material that
accumulates on the jet head in 2D
\citep{Lazzati2010,Mizuta2013,Gottlieb2018}, but disappears in
3D. Another important aspect is whether the hollow core angular structure
found in 2D survives against potential 3D jet instabilities like the kink
\citep[e.g.][]{Moll2008,McKinney2009,Mizuno2012}.

From the initial setup, the MRI develops inside the torus and drives
accretion onto the black hole (more details in the Appendix \ref{appen}).
As matter plunges onto the black hole it contributes to the accumulation
of magnetic flux over the event horizon and a magnetically dominated
funnel starts to emerge. At the same time, magnetic pressure drives an
MHD wind from the outer layers of the torus. Subsequently, an MHD jet is
launched along the axis of rotation of the black hole while it is
confined by the torus and the surrounding high rest-mass density
material. After launch the jet starts to accelerate and finally breaks
out of the merger ejecta. At such times, the outflow can be clearly
defined by two distinguishable components: the relativistic jet with bulk
Lorentz factor $\Gamma\gg1$ and the MHD wind (ejecta) with velocities
that span from ${\rm 0.1\,-\, 0.9\ c}$. Most of this mass in the latter
component is sub-relativistic and only a tail of this acquires high
velocities. Such a fast-moving tail component has been indicated and
discussed in BNS merger simulations \citep{Hotokezaka2013,
  Hotokezaka2018, Bauswein2013b, Radice2018a}. During the jet formation
and acceleration we have not detected the development of a kink
instability; this may be due to the strong poloidal magnetic field
employed in the initial configuration. It is possible, however, that such
instability will develop on timescales longer than those explored with
our simulations and if a strong toroidal component has
developed. Indeed, following \citet{Appl2000}, we can obtain
  approximate estimates of the growth-time of the kink instability
  $\tau_{\rm kink}$ at different times during the simulation, finding
  that $\tau_{\rm kink} \sim 39\,{\rm ms}$ after $15\,{\rm ms}$ in the
  evolution and that that $\tau_{\rm kink} \sim 220\,{\rm ms}$ after
  $20\,{\rm ms}$. Clearly, these timescales are always larger than the
  simulation time ($\simeq 20\,{\rm ms}$).

\begin{figure*}
  \centering
  \includegraphics[width=0.77\textwidth]{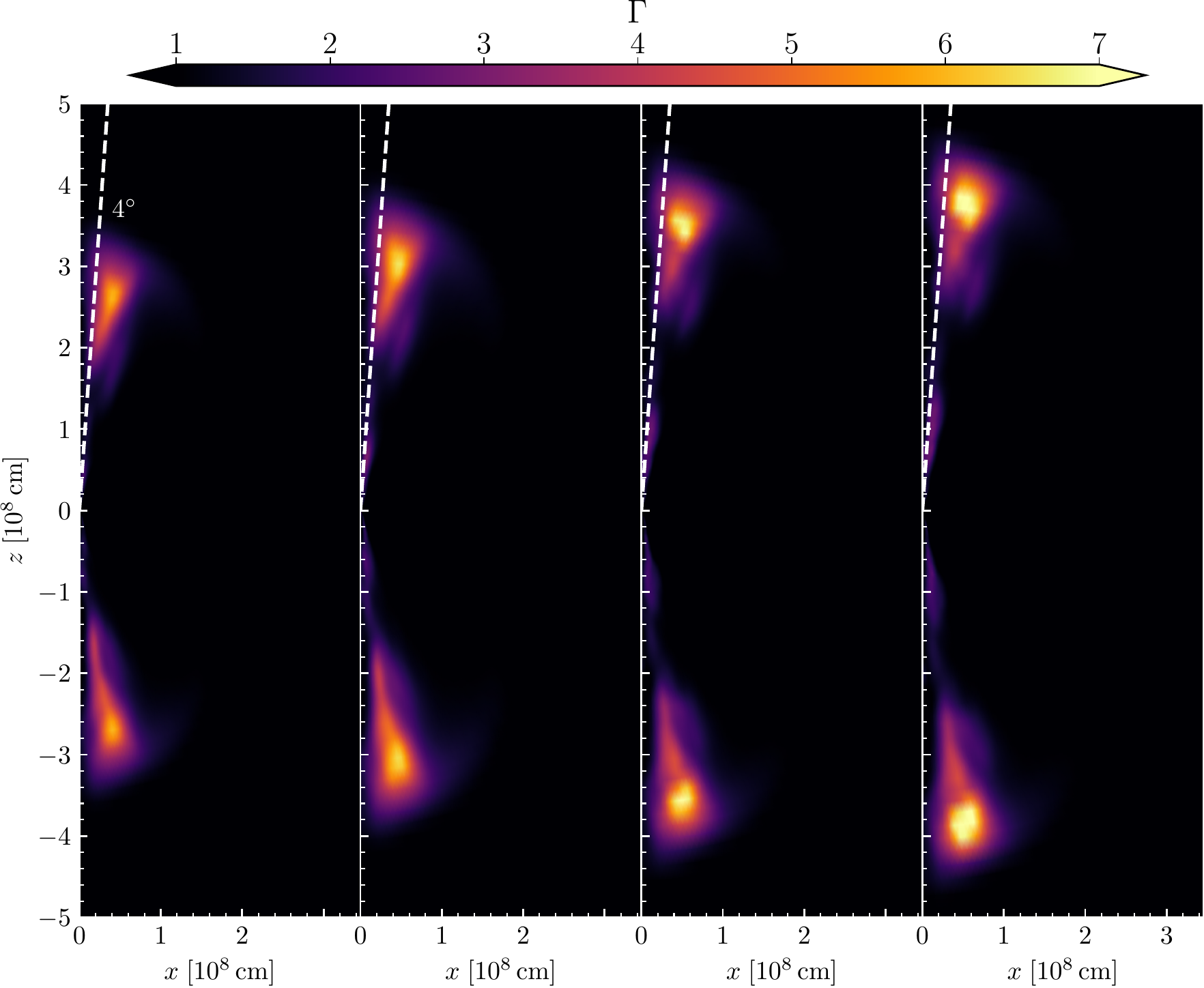}
  \hskip 1.0cm
  \includegraphics[width=0.77\textwidth]{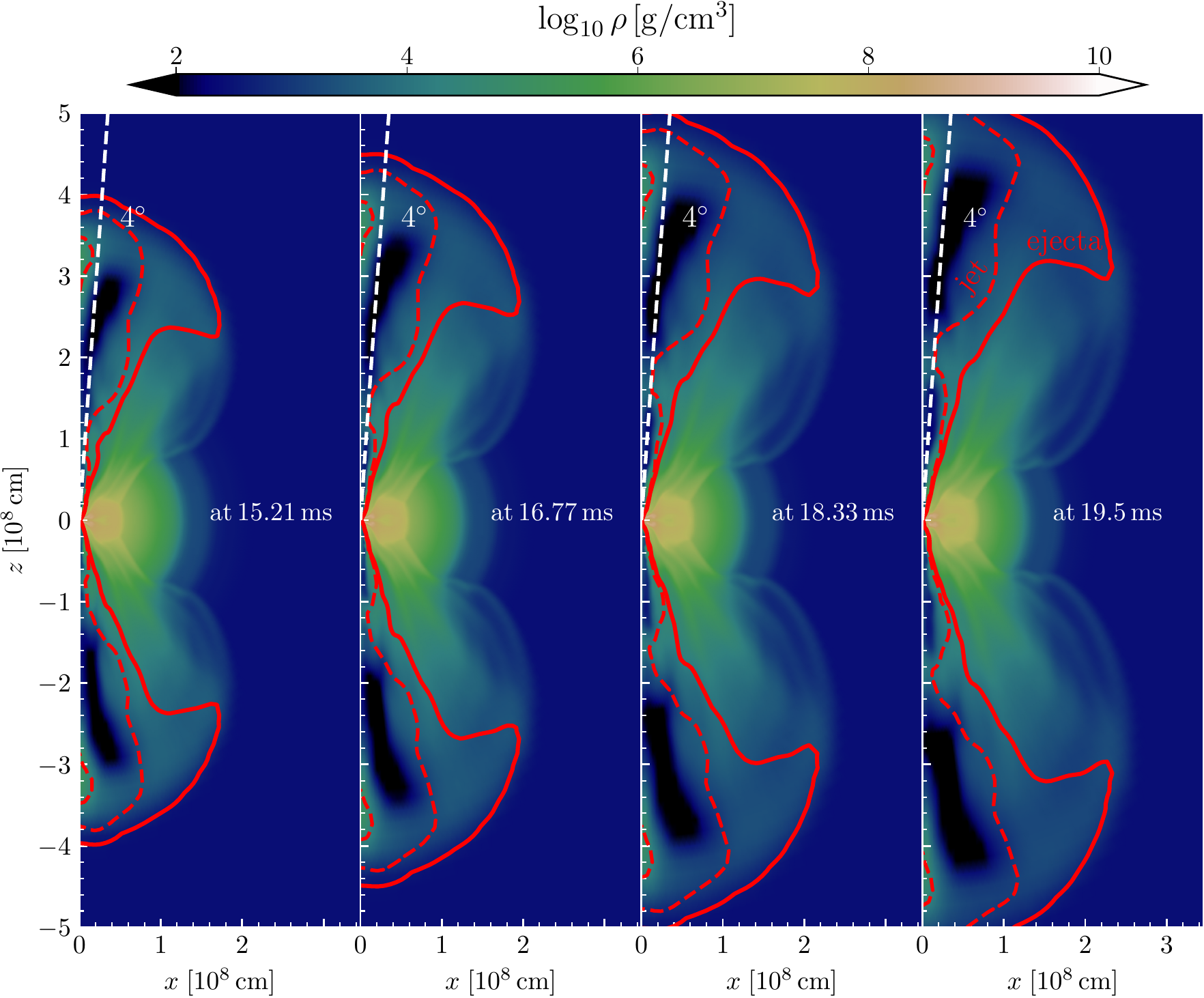}
  \caption{Lorentz factor (upper panels) and rest-mass density (lower
    panels) distribution for four consecutive snapshots azimuthally
    averaged. A cone with opening angle of $4^{\circ}$ is indicated with
    a dashed white line to highlight the slow moving core. On the lower
    panels the red solid lines denote the contour of $hu_t= -1$
    (gravitationally unbound), so that matter above such line is
    considered as merger ejecta, the red dashed lines denote the contour
    of $hu_t= -2.6$, inside this line matter has Lorentz factor
    $\Gamma>2$, and is considered to be in the jet.}
  \label{fig:MHDjet}
\end{figure*}
\begin{figure*}
  \centering
  \includegraphics[width=0.77\textwidth]{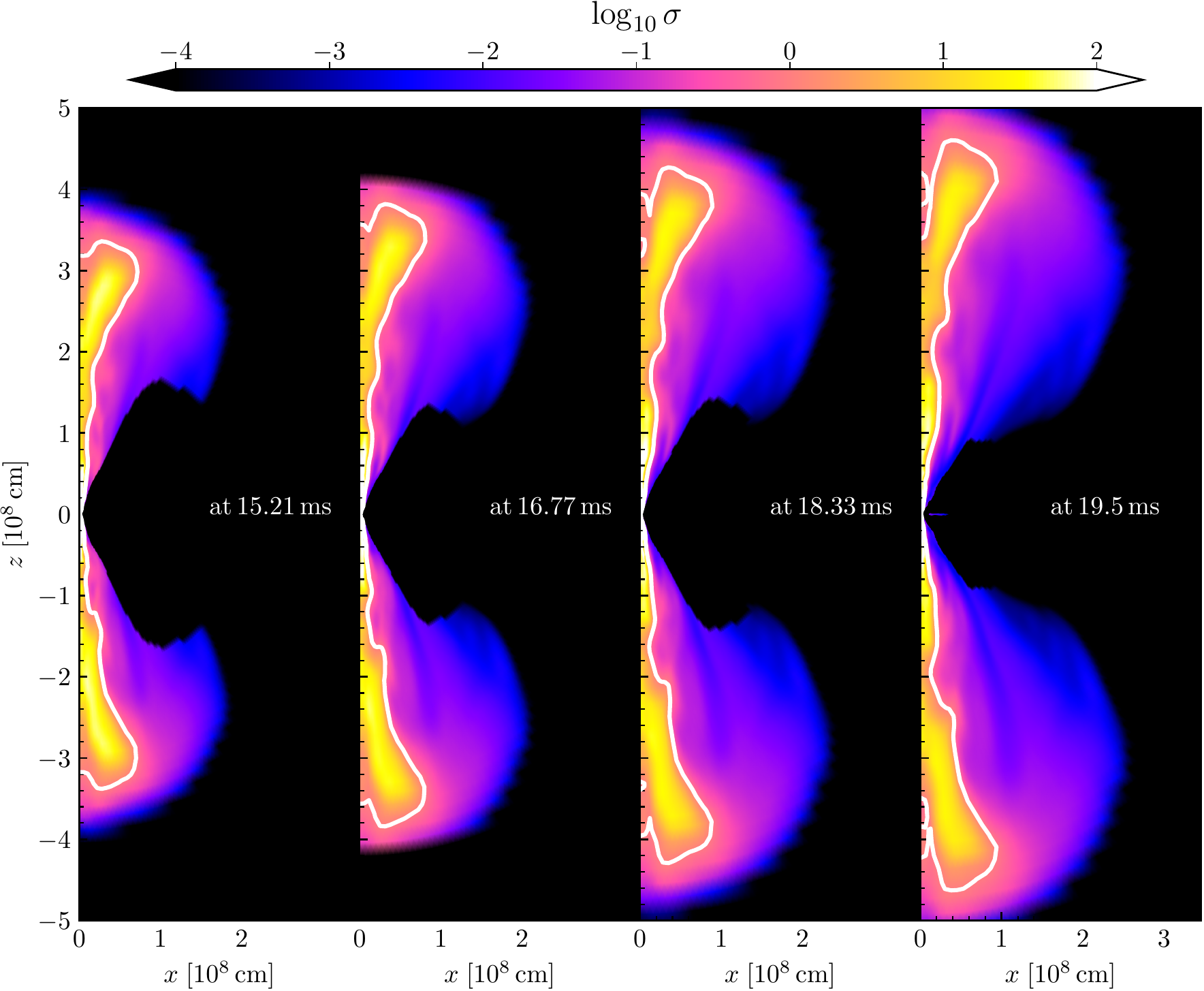}
  \caption{Magnetisation $\sigma$ for four consecutive snapshots
    azimuthally averaged. The white solid lines denote the contour of
    $\sigma= 1$, the part of the outflow enclosed by the contour is
    highly magnetized and will be experience efficient magnetic
    acceleration.  }
  \label{fig:sigma}
\end{figure*}
\begin{figure*}
  \centering
  \includegraphics[width=0.30\textwidth]{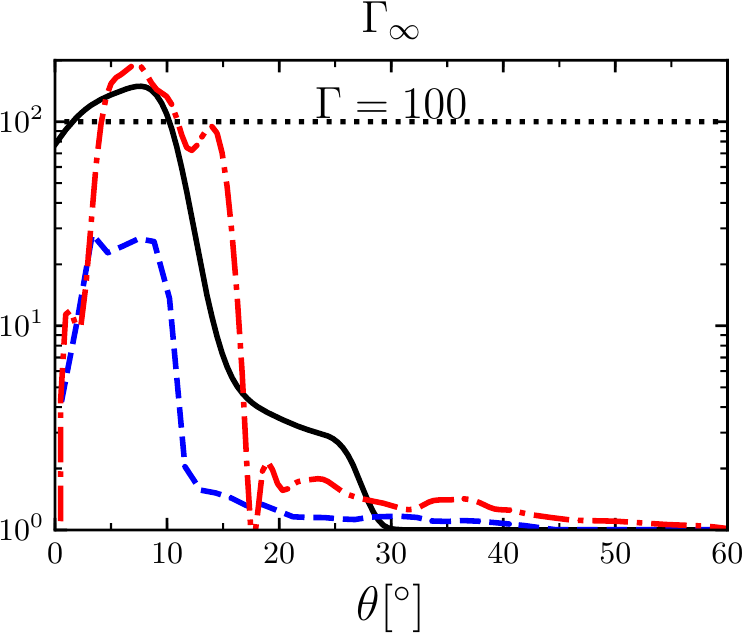}
  \hskip 0.1cm
  \includegraphics[width=0.30\textwidth]{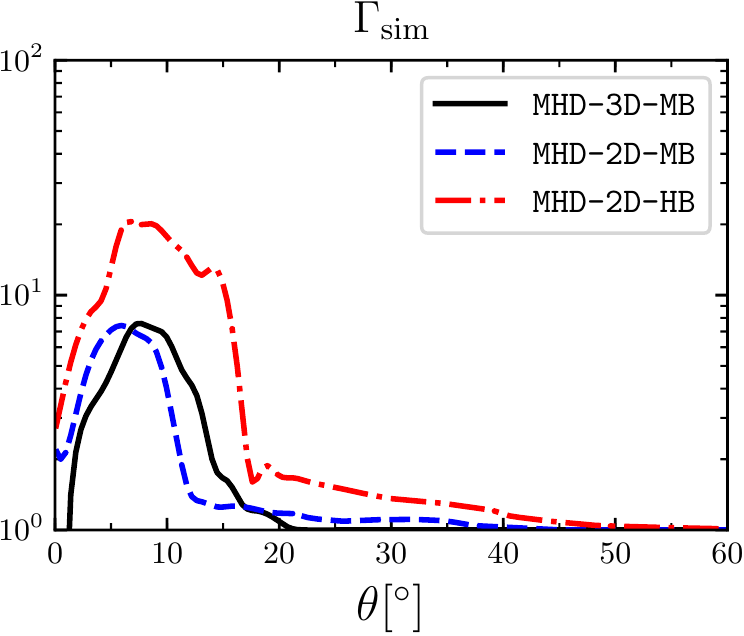}
  \hskip 0.1cm
  \includegraphics[width=0.30\textwidth]{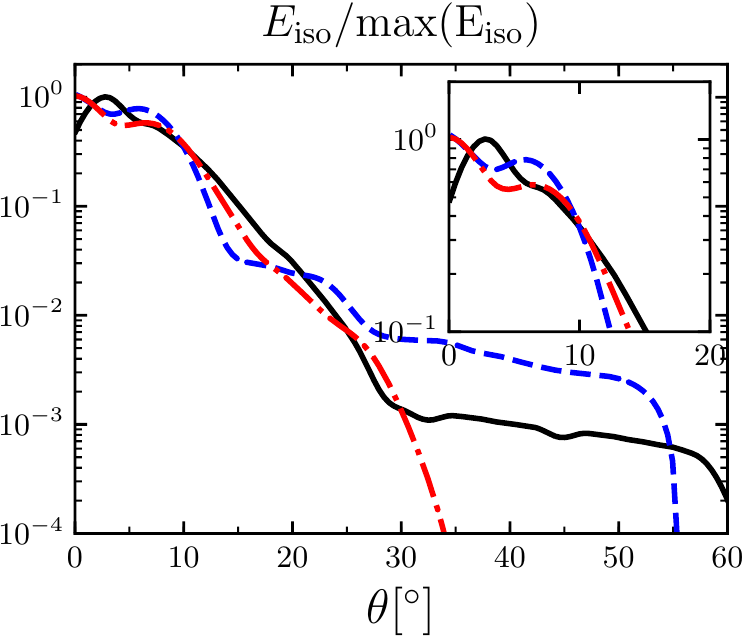}
  \vskip 0.2cm
  \includegraphics[width=0.90\textwidth]{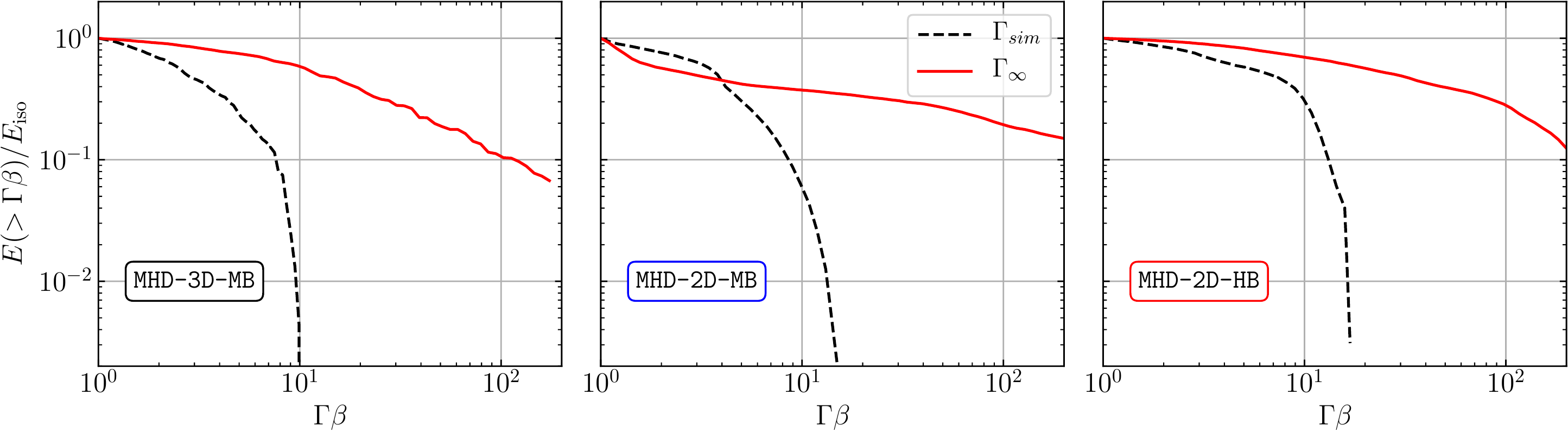}
  \caption{Left upper panel: Terminal Lorentz factor for the three models
    considered. Middle upper panel: Lorentz factor of the simulation,
    Right upper panel: the normalized energy 
    distribution across $\theta$, the in-box
    plot is a zoom in at small angles. Model \texttt{MHD-3D-HB} is shown
    in black solid lines, \texttt{MHD-2D-HB} in red dashed-doted lines
    and \texttt{MHD-2D-MB} in blue dashed lines. Lower panels: Normalized
    energy distributions $E(>\Gamma\beta)/E_{\rm iso}$ for the three
    models that is approximately obtained for all polar angles.  The
    black dashed lines corresponds to the distribution computed using the
    Lorentz factor from the simulation output, where as the red solid
    lines are computed using the Terminal Lorentz factor.  }
 \label{fig:ang}
\end{figure*}
\begin{figure*}
  \centering
  \includegraphics[width=0.30\textwidth]{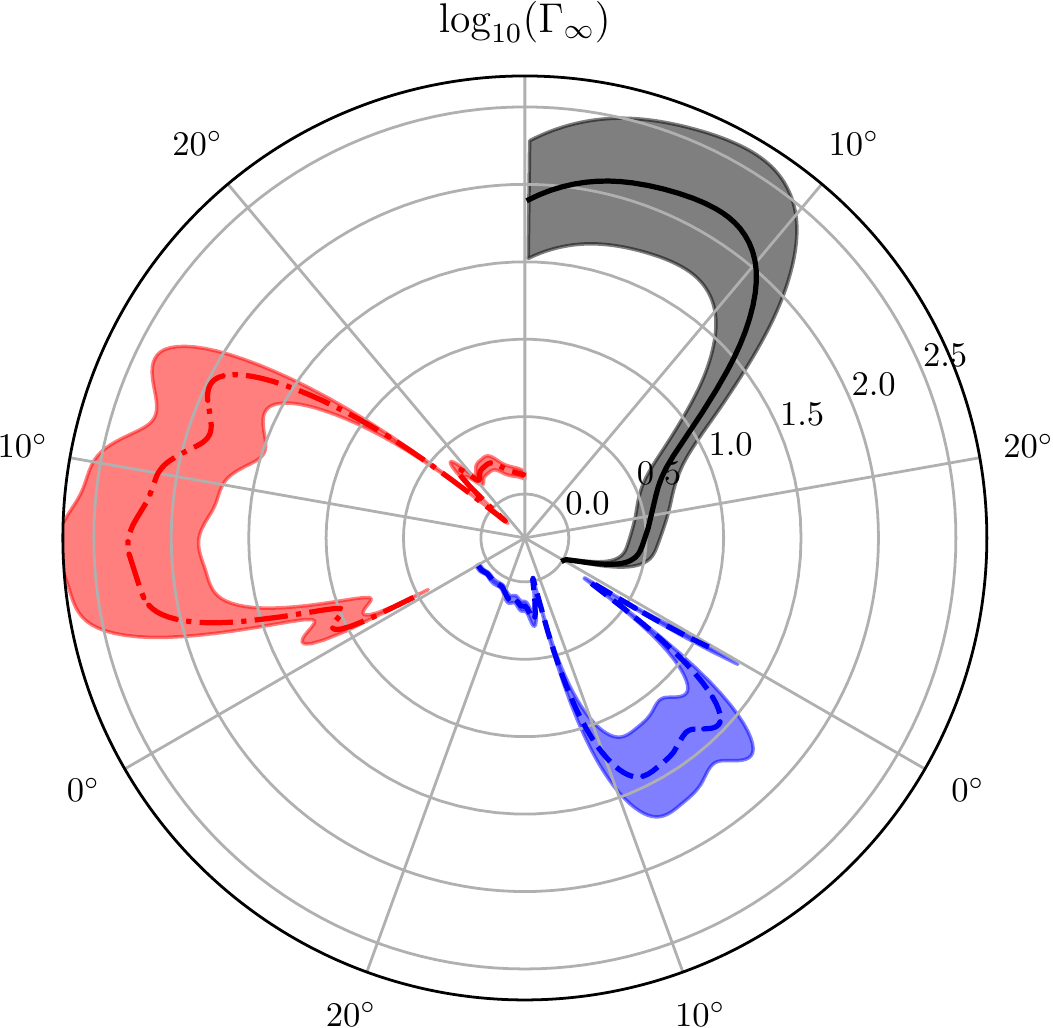}
  \hskip 0.1cm
  \includegraphics[width=0.30\textwidth]{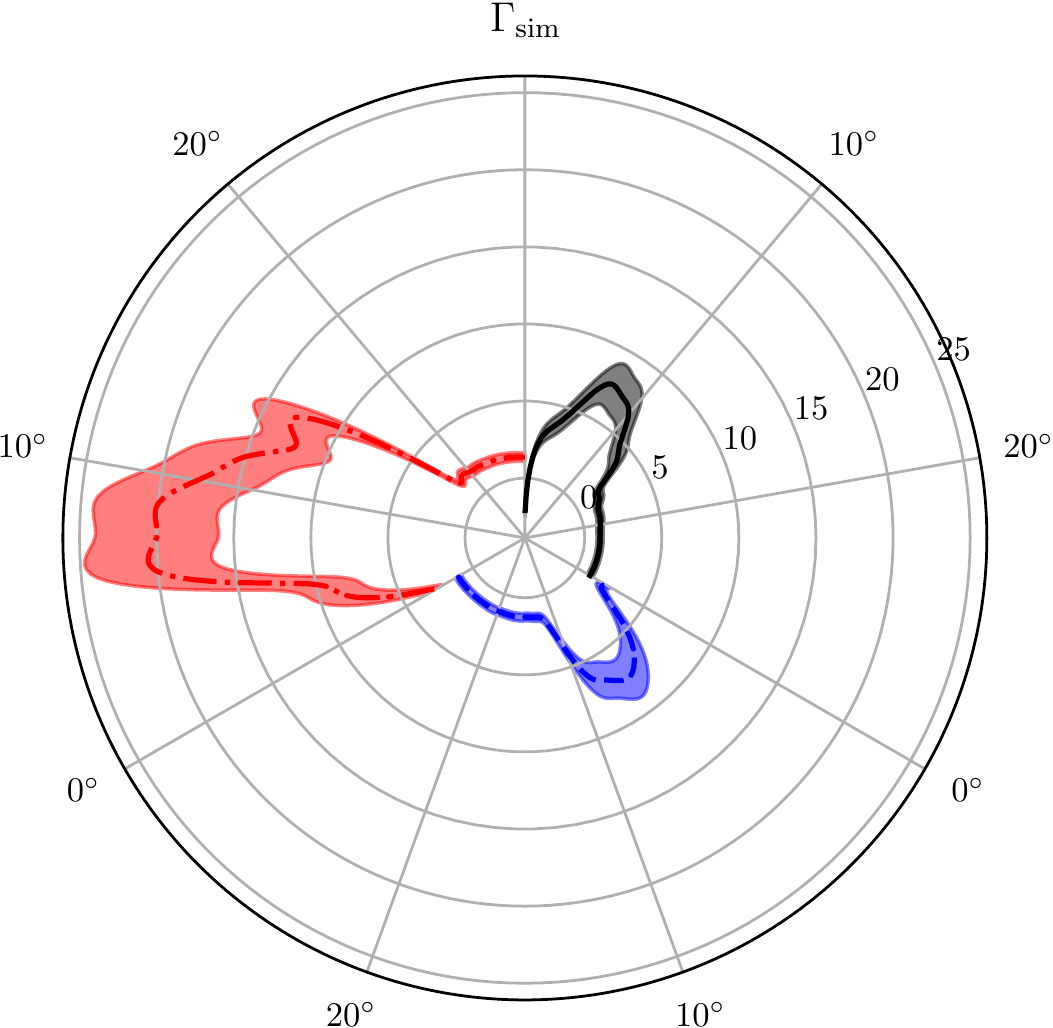}
  \hskip 0.1cm
  \includegraphics[width=0.30\textwidth]{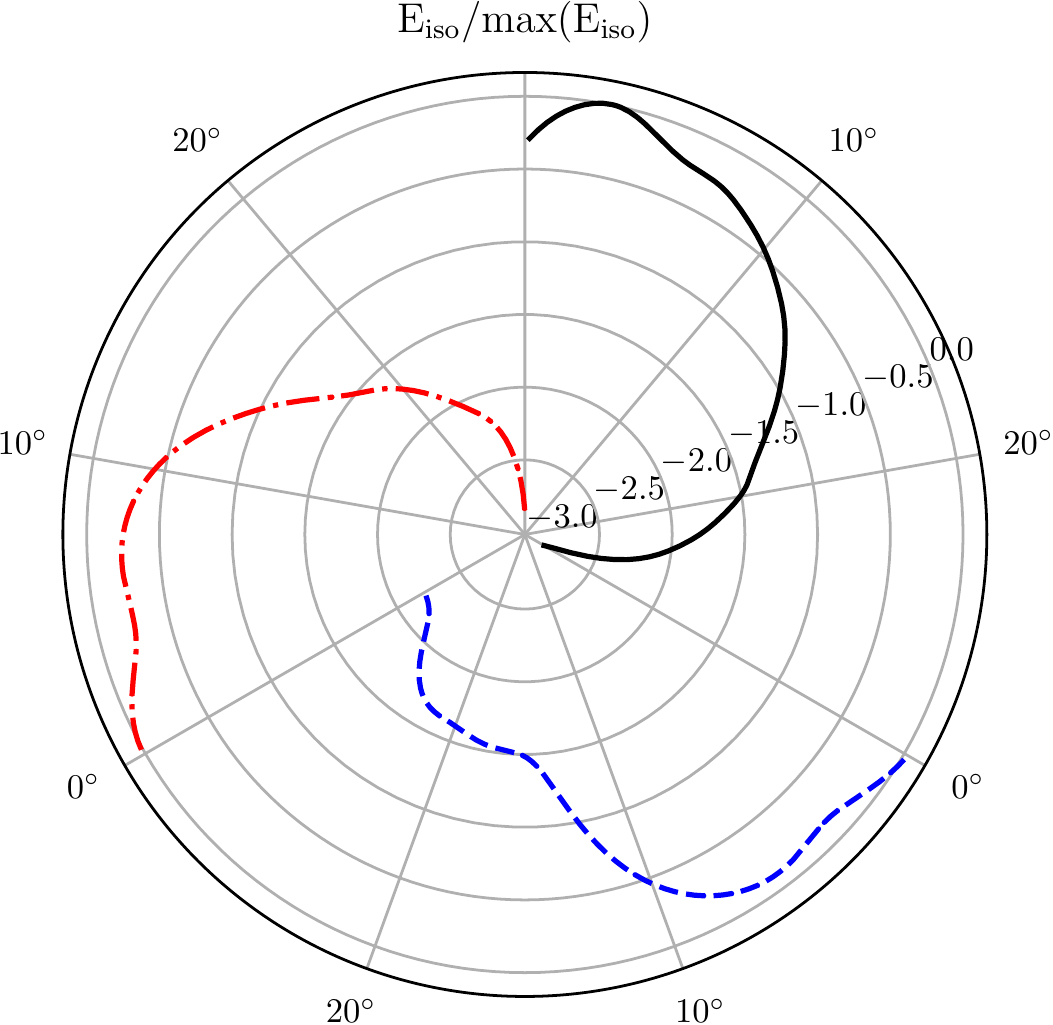}
  \caption{Left panel: Polar plot of the logarithm of the terminal
    Lorentz factor for the three models considered. Middle panel: Polar
    plot of the Lorentz factor from the simulation, Right panel: Polar
    plot for the energy distribution across $\theta$. Model
    \texttt{MHD-3D-HB} is shown in black solid lines, \texttt{MHD-2D-HB}
    in red dashed-doted lines and \texttt{MHD-2D-MB} in blue dashed
    lines. In the left and middle panels the shaded area corresponds to
    the 1-$\sigma$ variance. }
 \label{fig:polar}
\end{figure*}
In Fig. \ref{fig:MHDjet} we present the resulting structure of the MHD
jet post break-out from the matter distribution.
Both panels show four consecutive times for the development of the shape
of the jet and clear distinction of the MHD ejecta component. Simulation
data are averaged in the azimuthal $\phi$ direction. In the upper panel
we show the Lorentz factor of the outflow, where it is clearly seen that
the jet posses a slow moving core within an angle of $\theta_{\rm
  core}\approx4^{\circ}$, this confirms our previous results from 2D
simulations. More specifically, the relativistic jet is confined to polar
angles $\theta_{\rm core}\lesssim\theta\lesssim\theta_{\rm jet}$, where
$\theta_{\rm jet} \approx 10^{\circ}-15^{\circ}$. The high end of
$\theta_{\rm jet}$ range is acquired as the flow advances in radius
$\approx 4\times 10^8\, {\rm cm}$ (left-most plot of the upper panel in
Fig. \ref{fig:MHDjet}). Such an opening angle is also inferred from
numerical-relativity simulations where the only the starting point for
the launching of such a jet is reached \citep{Rezzolla:2011, Kiuchi2014,
  Dionysopoulou2015, Kawamura2016, Ruiz2016}, and there is a need to
combine such results with what we simulate here \citep{Nathanail2018b}.

In the lower panels of Fig. \ref{fig:MHDjet} we present the rest-mass
density distribution as the jet breaks out. In order to quantify how much
matter becomes unbound, we use the Bernoulli criterion and assume a fluid
element to be unbound if it has $hu_t\leq -1$, where $h$ is the specific
enthalpy of the fluid \citep{Rezzolla_book:2013}. The red solid line
indicates the contour of $hu_t=-1$ that marks the boundary of the unbound
mass, which means that all matter above the red solid line can be safely
considered to be part either of the jet component or the ejecta
component. We apply the Bernoulli criterion to measure the amount of
unbound matter that passes through a 2-sphere of $1, 200\,{\rm km}$ and
report it in the last two columns of Table \ref{tab:initial}, the actual
amount of ejected mass and the fraction of the ejected mass with respect
to the initial mass of the torus. The amount of the ejected mass for
model \texttt{MHD-3D-HB} is $26.1\%$ of the initial mass which is similar
to what found from the 2D simulations\footnote{In our previous study due
to the much larger evolution times from all the models reported there,
the ejected matter was extracted after passing through a 2-sphere of $4,
000\,{\rm km}$ \citep{Nathanail2020b}.}.

The red dashed line corresponds to a contour of $hu_t=-2.6$ (almost
identical to the contour of $\Gamma=2$ and of $\sigma=1$, see below). We
consider matter inside the red dashed line to be part of the jet whereas
outside of this to be part of the ejecta. The fast-moving tail of the
ejecta that reaches velocities of $\approx 0.87\, {\rm c}$ lies closely
outside the boundary of the jet. The outflow that breaks out from the
merger ejecta is highly magnetized with magnetisation, the ratio of
magnetic field to cold matter enthalpy densities, $\sigma= B^2/\rho c^2
>10$. This is seen in Figure \ref{fig:sigma}, where we azimuthally
average the magnetisation of the outflow and also denote the $\sigma=1$
contour by a white solid line.

At this stage it is expected that the jet has not fully undergone
complete MHD acceleration, thus we obtain both the Lorentz factor
straight from the simulation, $\Gamma_{\rm sim}$, is the maximum Lorentz
factor at a certain angle, as well as the terminal Lorentz factor,
$\Gamma_\infty$, expected after MHD acceleration. It has been shown
semi-analytically and numerically that a confined relativistic outflow
highly magnetized (Fig. \ref{fig:sigma}), can be efficiently accelerated
and attain a Lorentz factor more than $50\%\, {\rm \mu}$, depending on
the details of the solution \citep{Vlahakis2003a,Vlahakis2003b,
  Tchekhovskoy2008,Komissarov:2009,Lyubarsky2010}, where ${\rm \mu}$ is
the ratio of the total energy flux to mass flux. We calculate
$\Gamma_\infty$ for the part of the outflow that passes radially
$>2\times 10^8 \, {\rm cm}$ by averaging the energy-weighted ${\rm \mu}$
over $\phi$ and time $t$ \citep{Kathirgamaraju2019}:
\begin{align}
 \Gamma_{\infty}= \frac{\int \mu T_t^r d\phi dt}{\int T_t^r d\phi dt}\,,   
\label{eq:gamterm}
\end{align}
where $ T_t^r$ is the radial energy flux that comes from the
stress-energy tensor (subtracting rest mass energy). The above estimation
gives a measure of how a magnetically dominated outflow efficiently
converts its magnetic energy (although the flow could also have other
sources of acceleration, \eg thermal energy) to kinetic energy and
accelerates the flow \citep{Bromberg2016, Lyubarsky2010b,
  Sapountzis2013}.  However, a consistency check is needed to verify if
this acceleration is possible before the time that the outflow has
started decelerating. For an outflow of an initial Lorentz factor of a
few $\Gamma \gg 1$, similar to what we have from our simulations, it has
been shown that the acceleration region needs to be at least 4 orders of
magnitude in distance, thus taking our initial point the jet break-out,
which is at around $\approx 10^8 \, {\rm cm}$ in our simulation, then the
flow has efficiently accelerated at a distance of $\approx 10^{12} \,
{\rm cm}$.  For ultra-relativistic flow ($\Gamma_\infty\gg1$)
deceleration occurs when the swept up mass, $M_{\rm sw} =
(4\pi/3)R^3m_pn_{_{\rm ISM}}$ exceeds $M_{\rm
  ej}(\theta)/\Gamma_\infty(\theta) = E_{\rm k,iso}
(\theta)/\Gamma_\infty^2(\theta)c^2$, where $M_{\rm ej}(\theta)$ and
$E_{\rm k,iso}(\theta)$ are the entrained mass and the
isotropic-equivalent kinetic energy of the outflow at polar angle
$\theta$. From this, we find the deceleration radius
\begin{align}
 r_d(\theta) =& \left(\frac{3E_{\rm k,iso}(\theta)}{4\pi m_pn_{_{\rm ISM}}c^2\Gamma_\infty^2(\theta)}\right)^{1/3}  \nonumber \\
 =& 4.64\times 10^{16}\left(\frac{E_{\rm iso}}{4\times10^{51}\, 
 {\rm erg}}\right)^{1/3} 
 \!\!\times\!\! \left(\frac{\Gamma_\infty}{100}\right)^{-2/3}
 \left(\frac{n_{_{\rm ISM}}}{10^{-2}}\right)^{-1/3}\, \, {\rm cm}\,, \nonumber
\label{eq:rd}
\end{align}
\begin{figure*}
  \center
  \includegraphics[width=0.44\textwidth]{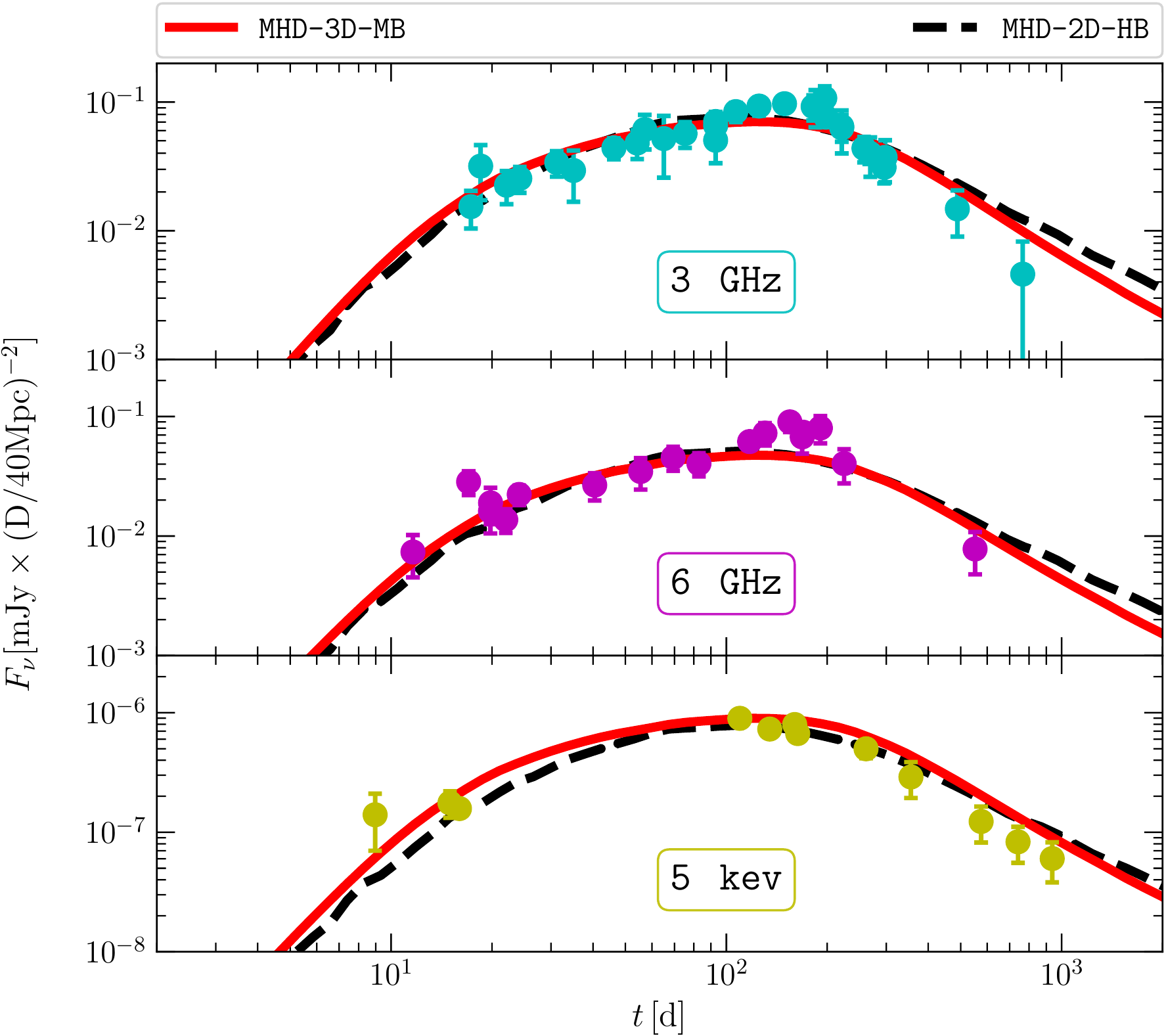}
  \hspace{0.5cm}
  \includegraphics[width=0.49\textwidth]{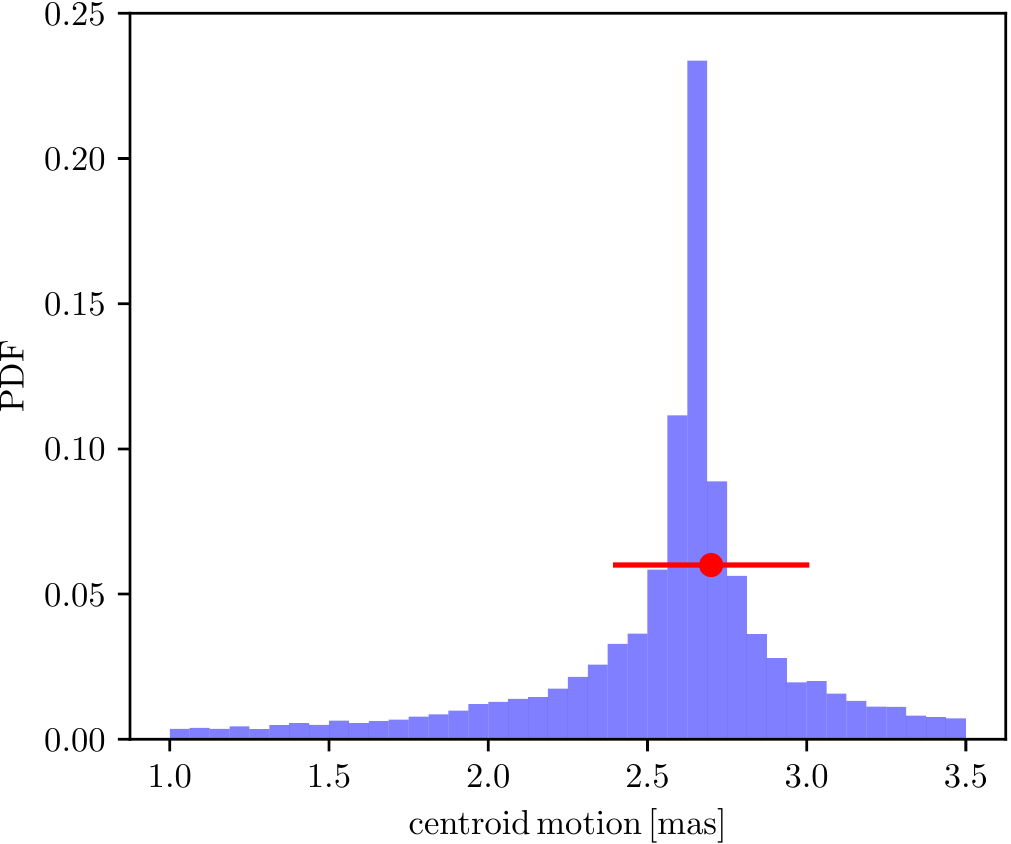} 
  \caption{Left panel: Best-fit lightcurves of models \texttt{MHD-3D-HB}
    (red line) and \texttt{MHD-2D-HB} (dashed blue line) for the
    broadband afterglow observations of GRB170817A. The corner plot for
    the analysis to find the best-fit parameters from the genetic
    algorithm is shown in Fig. \ref{fig:corner}. Right panel: The
    distribution of the flux centroid motion for model
    \texttt{MHD-3D-MB}, the red dot with error bars correspond to the
    observed value for the flux centroid motion $2.7\pm 0.3\, {\rm mas}$
    \citep{Mooley2018b}. }
	\label{fig:aft}
\end{figure*}
where for the final expression we use fiducial values that are relevant
for the relativistic jet. Note that a lower Lorentz factor yields a
larger $r_d$, thus we are confident that any acceleration occurs before
the deceleration stage and use the asymptotic Lorentz factor
$\Gamma_{\infty}$, for the calculation of the afterglow.

In the upper left and middle panels of Fig. \ref{fig:ang} we show the 
distribution of the Lorentz factor, for the three models considered in Table
\ref{tab:initial}, that is averaged azimuthally and extracted from the
part of the outflow that reaches a radial distance $>2\times 10^8\, {\rm
  cm}$. In the upper left panel we present the terminal Lorentz factor
$\Gamma_\infty$, computed from Eq. \eqref{eq:gamterm} and in the middle
panel the average Lorentz factor $\Gamma_{\rm sim}$. We find that
$\Gamma_{\rm sim}$ obtained from the 3D simulation seems to be closer to
that obtained from the 2D case that has a medium magnetic energy (model
\texttt{MHD-2D-MB}). This, however, does not hold when calculating the
terminal Lorentz factor as shown on the left panel of the Figure. Due to
the available magnetic energy for acceleration in model
\texttt{MHD-3D-HB}, its distribution of the terminal Lorentz factor is
more similar to the high magnetic model \texttt{MHD-2D-HB}. In the
\texttt{MHD-3D-HB} model the distribution of Lorentz factor (both
simulation and terminal) peak at an angle $\theta\approx8^{\circ}$.

In the rightmost upper panel of Fig. \ref{fig:ang} we plot the 
isotropic energy of the outflow normalized with the maximum energy, due 
to the fact that the medium magnetic field, \texttt{MHD-2D-MB} has a 
smaller energy and it would be difficult for comparison. The maximum 
isotropic energies for models \texttt{MHD-3D-HB}, \texttt{MHD-2D-HB} and 
\texttt{MHD-2D-MB} are $E= 2.7\times 10^{51}\,{\rm erg}$, $3.6\times 
10^{50}\,{\rm erg}$, and $2.5\times 10^{49}\,{\rm erg}$ respectively. In 
the 3D model \texttt{MHD-3D-HB}, the energy peaks at $\theta\approx
3^{\circ}$ and decreases for smaller angles. The main difference from 
the high magnetic model \texttt{MHD-2D-HB}, is that the 3D model has an 
extended energy distribution to larger angles till $\theta\lesssim
60^{\circ}$. The angular distribution of energy is similar to model 
\texttt{MHD-2D-MB}, but for this model the normalization is two orders 
of magnitude smaller.

In the lower panels of Fig. \ref{fig:ang} we show the distribution
of energy above a certain value of $\Gamma\beta$, i.e., $E(> \Gamma\beta)$,
as a function $\Gamma\beta$, normalized by $E_{\rm iso}(\theta)$. This
energy distribution profile is approximately obtained at all polar
angles. For all three models the black dashed line corresponds to the
calculation performed using $\Gamma_{\rm sim}$, whereas the red solid
line corresponds to the calculation using $\Gamma_\infty$. The energy
dependence, $E(> \Gamma\beta)\propto(\Gamma\beta)^{-s}$ with $s=4-5$,
when using the simulation Lorentz factor is similar to the one reported
in our previous work \citep{Nathanail2020b}. However, assuming that
acceleration is smooth and completes before any deceleration has started
(as was discussed earlier) and using the terminal Lorentz factor, this
dependence changes and the fast moving part of the outflow has a large
amount of energy. The energy reported in Fig. \ref{fig:ang}, is measured
after the jet has broken out from the merger ejecta, i.e., $t = 10\, {\rm
  ms}$.

Note that in our MHD simulations, a cut-off of $\Gamma=20$ is set, in
order to avoid parts of the outflow where the accuracy of the numerical
method is reduced because of the large Lorentz factors attained.
\section{Afterglow emission from the jet}
\label{sec:after}
\subsection{Description of the emission model}
\label{sec:rad}

We model the afterglow emission as synchrotron radiation from
shock-heated power-law electrons. The interaction of the relativistic
outflow with the surrounding ISM produces a forward shock that propagates
into the ISM and accelerates electrons into a power-law energy
distribution, $n_e(\gamma_e)\propto\gamma_e^{-p}$, where $n_e$ is the
number density and $\gamma_e$ is the Lorentz factor of the electrons. For
the slope of the distribution, we employ the value $p=2.138$, which was
recently updated from the panchromatic analysis of the afterglow of
GRB170817A \citep{Makhathini2020}. We further assume the standard
afterglow shock microphysics where a fraction $\epsilon_e$ and
$\epsilon_B$ of the total internal energy behind the shock is given to
electrons and magnetic field, respectively \citep{Sari1998}.

\begin{figure*}
  \includegraphics[width=0.93\textwidth]{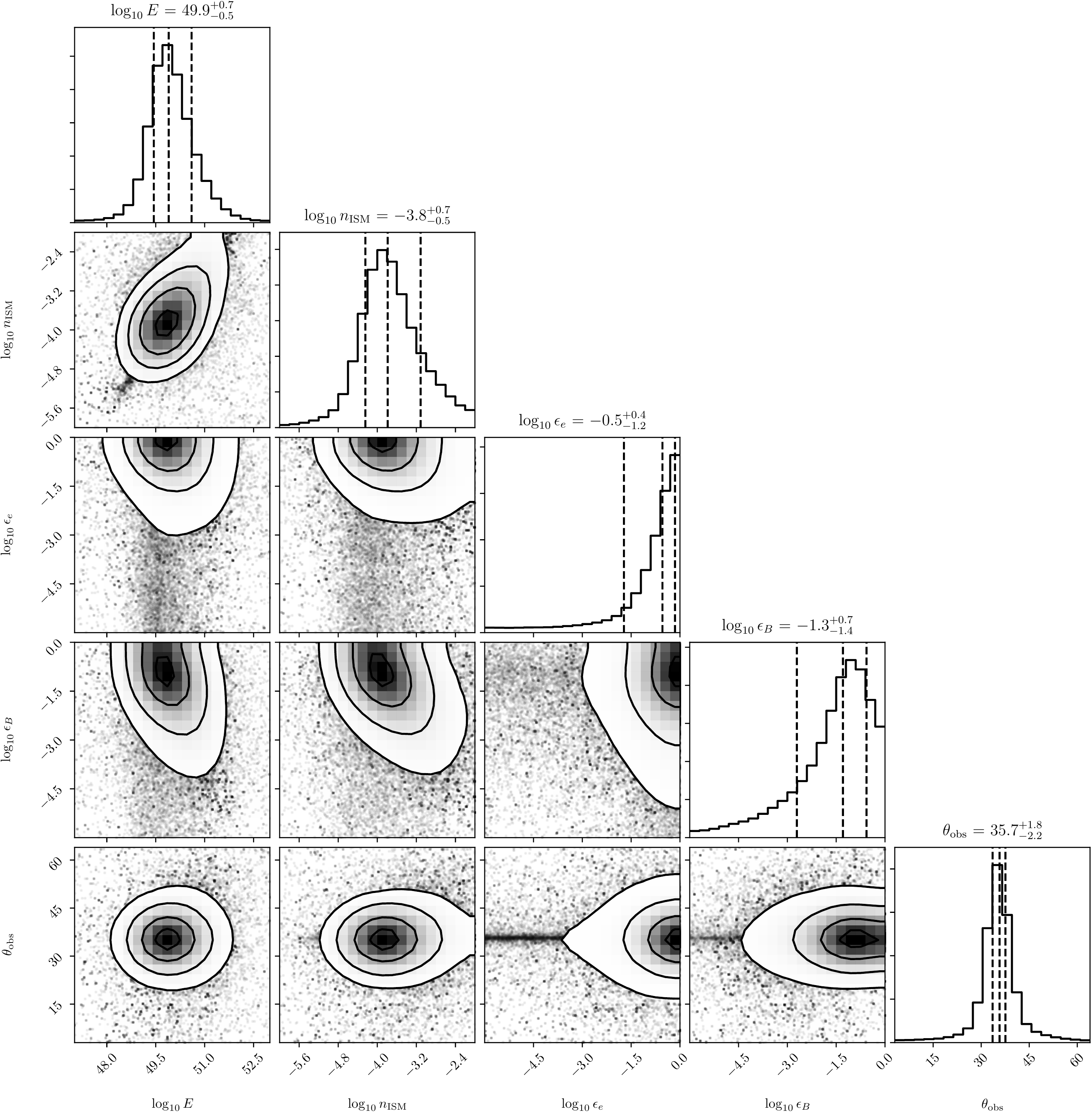}
  \caption{Corner plot from the analysis for the parameters of the
    best-fit lightcurves for model \texttt{MHD-3D-MB}, shown in
    Fig. \ref{fig:aft}. 1-$\sigma$ uncertainties shown in the titles, are
    computed as the 16th and 84th percentiles of the one-dimensional
    posterior distribution for each parameter. The 3-$\sigma$
    uncertainties for the intrinsic parameters are $\log_{10}E
    =49.9^{+1.6}_{-1.2}\, {\rm erg}$, $\log_{10} n_{_{\rm ISM}}=
    -3.8^{+1.4}_{-1.1} \, {\rm g/cm^3}$ and $\theta_{\rm obs}=
    35.7^{\circ \,\,+15.3}_{\quad-17.}$.  }
  \label{fig:corner}
\end{figure*}
To compute the afterglow lightcurves we use the angular distributions of
the Lorentz factor and of the energy profile (\cf
Fig.~\ref{fig:ang}). Along the $\theta$ direction, we use $200$ angles
uniformly distributed to acquire the angular structure of the outflow,
which then yields the initial Lorentz factor
$\Gamma_0(\theta)=\Gamma_\infty(\theta)$ and the isotropic-equivalent
energy $E_{\rm iso}(\theta)$ (after MHD acceleration) of the flow. We
calculate the afterglow lightcurve using a semi-analytic formalism 
where we 
treat the flow using locally spherical dynamics for which each part of
the flow expands as if it is part of a spherical flow
\citep{Granot1999,Gill2018}.

For our study, we employ the most recent afterglow data, \ie
$t\lesssim940\,{\rm d}$ post merger consisting of X-ray emission ,
optical emission and VLA radio observations
\citep{Margutti2017,Margutti2018,Alexander2017,Alexander2018,Hallinan2017,
  Mooley2018,Mooley2018c,Dobie2018,Nynka2018,Troja2018,Troja2019,
  Lamb2019,Fong2019,Hajela2019,Makhathini2020}. To perform the fit, we
use five free parameters: the energy of the burst $E$, and the
circum-merger density $n_{_{\rm ISM}}$, the shock microphysical
parameters $\epsilon_e$ and $\epsilon_B$, and the observer angle
$\theta_{\rm obs}$. The model parameters outnumber the available
constraints from the data and thus the parameter space is degenerate
\citep{Gill2019b}. The shape of the afterglow lightcurve near the peak
can only constrain the ratio $\theta_{\rm obs}/\theta_j$, where
$\theta_j$ is the opening angle of the jet if the flow has sharp edges
(for an angular structured jet this would correspond to the angular size
of the relativistic core) \citep{Nakar2020}. An additional constraint is
obtained from the VLBI proper motion of the flux centroid
\citep{Mooley2018b}.

To find the best-fit parameters we use a genetic algorithm to optimized
the parameter selection and minimise the reduced $\chi_{\nu}^2$
\citep{Fromm2019,Nathanail2020b}. The fitting procedure for the
lightcurve is applied to the full uniform dataset reported in
\citet{Makhathini2020}. Rather than fitting the observed flux in
different energy bands, we first extrapolate all the observations to a
given energy, \eg $3\,$GHz, and then carry out the fit to the
lightcurve. This procedure is feasible since the spectrum of the
broadband afterglow is described by a single power law with flux density
$F_\nu\propto\nu^{-0.569}$ across all epochs
\citep[e.g.,][]{Hajela2019,Makhathini2020}. Simultaneously in the
fitting procedure, we require that the flux centroid motion matches the
observed values\footnote{To obtain the correct flux centroid motion
during the fitting procedure we minimise the quantity $\chi_{\nu}^2 +
w_{fc} \times (\chi^2_{fc} -1)$, where $\chi_{\nu}^2$ comes from the
lightcurve and $\chi^2_{fc}$ from the centroid fitting, the weighting to
flux centroid fitting is set to $w_{fc}=1-2$.}. Moreover, we assume a
prior of $n_{_{\rm ISM}}<0.01\, {\rm cm^{-3}}$, as obtained from
independent analysis of the host galaxy
\citep{Hallinan2017,Hajela2019,Makhathini2020}.

\begin{figure}
  \includegraphics[width=0.45\textwidth]{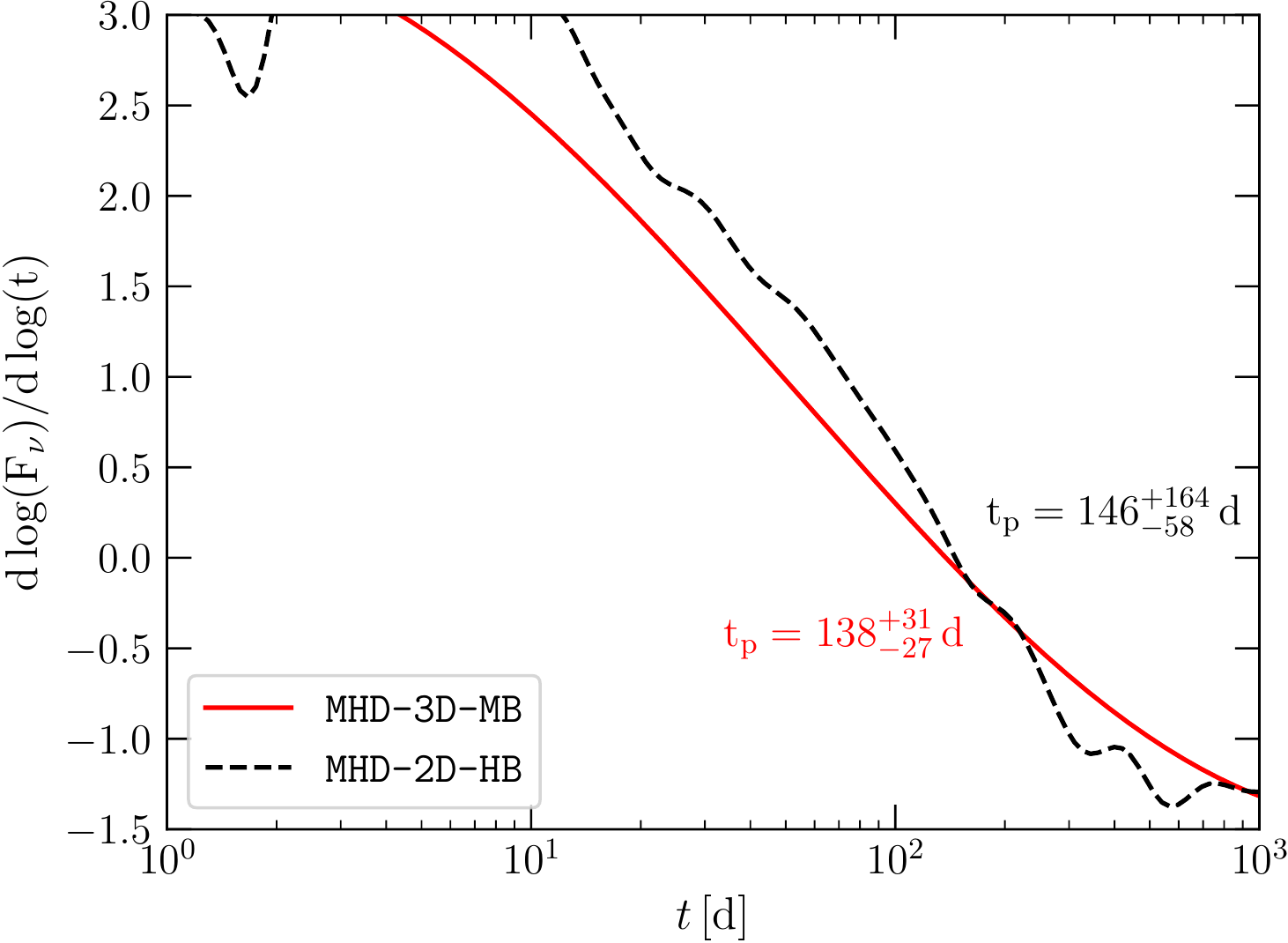}
  \caption{The slope of the best-fit lightcurves of models
    \texttt{MHD-3D-MB} (red solid) and \texttt{MHD-2D-HB} (dashed
    black). The peak of the lightcurve is reported in the plot with the
    $68\%$ confidence intervals. }
  \label{fig:slope}
\end{figure}

The best-fit lightcurves for models \texttt{MHD-3D-HB} (red line) and
\texttt{MHD-2D-HB} (dashed blue line) are shown in the left panel of
Fig. \ref{fig:aft}, whereas on the right panel we show the probability
distribution function for the flux centroid motion that comes from the
fitting procedure. The flux centroid was calculated using the
  same semi-analytic formalism used to calculate the afterglow
  lightcurve. In particular, here we follow the prescription given in
  \citet[][see their Eq.~(27)]{Gill2018} to first form an image of the
  afterglow emission on the plane of the sky and then calculate the flux
  weighted location of the centroid. Just like the calculation of the
  afterglow lightcurve, this procedure relied on the input of the angular
  structure of the flow from the numerical simulation, namely $E_{\rm
    iso}(\theta)$ and $\Gamma_\infty(\theta)$.  The results of the
analysis of the parameter space from the genetic algorithm is shown in
the corner plot of Fig. \ref{fig:corner}.  Focusing on the lightcurve, we
compute the logarithmic derivative of the lightcurve in
Fig. \ref{fig:slope} for the two models \texttt{MHD-3D-MB} (red solid)
and \texttt{MHD-2D-HB} (dashed black). We find that the peak of the
lightcurve for both of them is consistent with observations. Overall, the
afterglow lightcurve of hollow-core jet are in good agreement with the
observational data from GRB170817A/GW170817
\citep{Nathanail2020b,Takahashi2019,Takahashi2020}.

%

\subsection{Comparison with previous models}

Several studies have produced lightcurves that fit the broad band
afterglow observations of GW170817 and analyze the parameter space. Most
importantly, through afterglow modelling, a better constraint for the
viewing angle can be obtained, which is poorly known from the GW signal
\citep{Abbott2017}. All methods for parameter estimation through
lightcurve fitting assume hydrodynamic outflows, and the general results
obtained hint to an opening angle of $\approx 3^{\circ}-6^{\circ}$ and a
observer angle of $15^{\circ}-35^{\circ}$. The largest observer angle
$\approx 35^{\circ}$ comes from the a 3D simulation of a structured model
from \citep{Lazzati2017c}, when fitting the panchromatic afterglow up to
$940\,{\rm d}$ post-merger \citep{Makhathini2020}. The lowest end of the
observer angle,$\approx 16^{\circ}$, comes from studies that fit also the
VLBI constraints \citep{Mooley2018b,Hotokezaka2019,Ghirlanda2019}.

\begin{figure*}
  \begin{center}
    \includegraphics[width=0.45\textwidth]{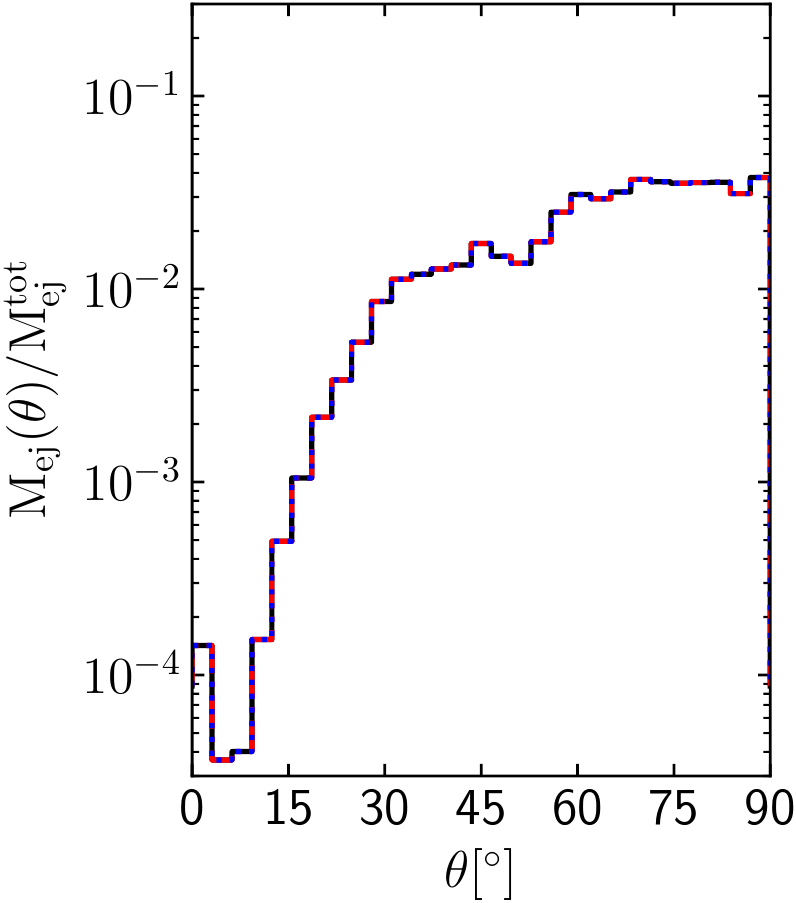}
    \vspace{1.0cm}
    \includegraphics[width=0.45\textwidth]{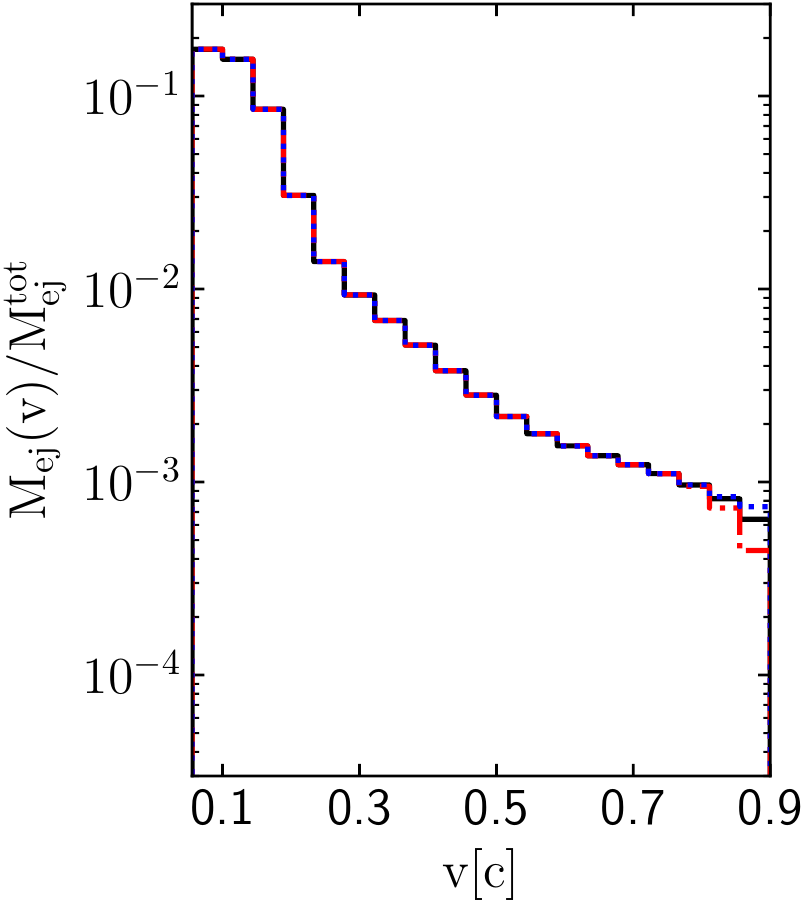}
  \end{center}
  \caption{Normalized mass distributions for the ejected mass (MHD wind)
    across the angle $\theta$ (left panel) and velocity (right panel).
    The total amount of ejected mass from the 3D simulation is
    $0.027\,M_{\odot}$, and it is defined as unbound matter with
    magnetisation $\sigma <1$ (blue dotted), different cut-offs are shown
    also, Lorentz factor $\Gamma<2$ (black solid), velocity $v<0.82$ (red
    dashed).}
  \label{fig:ej-distr}
\end{figure*}

Semi-analytical calculations for a Gaussian or a power-law jet with an
angular structure yield a observer angle of $\approx
20^{\circ}-27^{\circ}$
\citep{Resmi2018,Hotokezaka2019,Troja2019,Lamb2019, Ryan2020}. The only
studies that conclude to an observer angle of $> 30^{\circ}$ from their
parameter analysis are based on hydrodynamic simulations, which however
do not include the VLBI proper motion constraints
\citep{Lazzati2017c,Wu2019,Hajela2019}. A series of hydrodynamic
simulations were used from \citet{Mooley2018b} to fit both the lightcurve
and the proper motion and yielded an observer angle of $\approx
20^{\circ}$. Thus \citet{Makhathini2020} highlights the importance of the
use of the VLBI constraints together with the uniform afterglow dataset
to attain accurate estimates for the observer angle and other parameter
of the outflow.

In our study we use the uniform dataset and fit the lightcurve together
with the requirement for the flux centroid motion. From our analysis,
presented in Fig. \ref{fig:corner}, we obtain an observer angle of
$\theta_{\rm obs}= 35.7^{\circ \,\,+1.8}_{\phantom{\circ \,\,}-2.2}$,
which is the only one deduced from GRMHD simulations. It is interesting
that the analysis of the afterglow with hydrodynamic outflows that use
also the VLBI constraints, point to an observer angle
$\theta_{obs}\approx 16$ which is far different from what we report
here. The improved measurement of the observer angle from the GW is
$\theta_{\rm obs}= 29^{\circ \,\,+11}_{\quad-15}$
\citep{Abbott2018a}. Analysing the GW and combining a direct measurement
of the distance to the host galaxy of GW170817 yields an observer angle
$\theta_{\rm obs}=32^{\circ \,\,+10}_{\quad-13}$
\citep{Finstad2018}. Using broadband photometry of the kilonova AT2017gfo
associated with GW170817, results to independent constrains for the
observer angle $\theta_{\rm obs}=32.5^{\circ \,\,+11.7}_{\quad-9.7}$
\citep{Dhawan2020}. All the above constraints are compatible (and in good
agreement) with the MHD modelling presented in this paper.

\section{Afterglow emission from the ejecta (kilonova)}
\label{sec:eject}

The event GW170817 was followed by a kilonova detection which peaked in
the ultraviolet in a few hours and in the infrared few days post merger
\citep{Arcavi2017,Nicholl2017,Chornock2017,
  Cowperthwaite2017,Villar2017}. The modelling of this emission yielded a
huge amount of r-processed neutron rich mass $\approx 0.05\, M_{\odot}$
\citep{Kasliwal2017,Kasen2017} that was ejected during and after the
merger. This matter is expected also to produce non-thermal afterglow
emission similar to what was described in Section \ref{sec:after}. There
are studies in the literature that model this emission either for the
dynamical ejecta alone 

\citep{Hotokezaka2015,Radice2018a} or parametrize
them by prescribing a specific dependence on $\Gamma\beta$
\citep{Kathirgamaraju2019b,Hajela2019,Troja2020}.

\begin{figure}
  \includegraphics[width=0.45\textwidth]{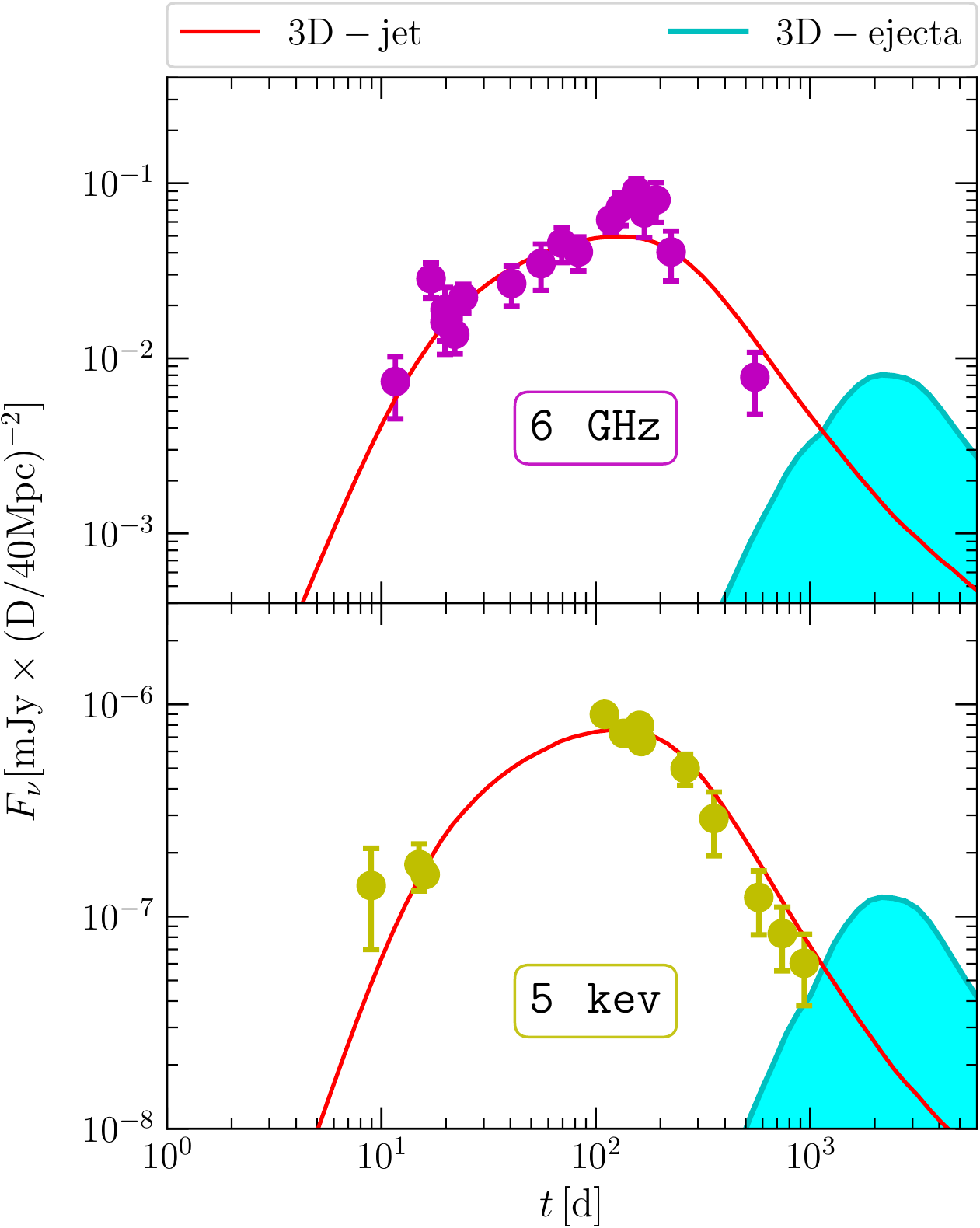}
  \caption{The jet component (red line) of model \texttt{MHD-3D-MB} is
    plotted against the broad band afterglow observations of GRB170817A,
    whereas the shaded cyan region corresponds to the allowed parameter
    space that the ejecta (kilonova) non-thermal afterglow may appear.}
\label{fig:eject-fut}
\end{figure}

The GRMHD outflow that we have described so far has two rather distinct
components: the relativistic jet confined in an angle of $\approx
15^{\circ}$ and a wide angle MHD wind that goes to very larger angles. In
this Section we extract from the 3D simulation the properties of this
ejected MHD wind and compute the distinct non-thermal afterglow that is
expected from it. This ejected MHD wind component is matter dominated
from the beginning, with $\sigma\ll1$ as seen in Fig. \ref{fig:sigma}, so
no MHD acceleration is expected. We need to note that by the way the
afterglow for the jet component was computed, the part of the flow that
moves with a Lorentz factor of $< 1.8$ (upper left panel of
Fig. \ref{fig:ang}), has a negligible effect on the computed lightcurve.
The ejected mass that we use for the non-thermal afterglow of the
kilonova, is defined as the unbound matter ($hu_t\leq -1$) that has
magnetisation $\sigma <1$ and thus is located outside of the jet. The
distribution of the ejected mass applying this cut-off is shown in
Fig. \ref{fig:ej-distr} across the angle (left plot) and the velocity
(right plot). The non-thermal afterglow emission from the ejected matter
(MHD wind), is calculated as discussed in the previous section but using
the distribution from Fig. \ref{fig:ej-distr}. In
Fig. \ref{fig:eject-fut} we use the latest observations in the X-rays at
$940\, {\rm d}$ post merger, to find the allowed parameter space that the
emission from the ejecta component may be observed in the future. In
order to find this allowed parameter space, we use the observer angle
$\theta_{\rm obs}=35.7^{\circ}$ (see Fig. \ref{fig:corner}). The energy
of the ejecta is known from the mass-velocity distribution. The ejected
mass from our 3D simulation is $0.027\,M_{\odot}$, and the total ejected
mass inferred from kilonova modelling is up to $0.07\,M_{\odot}$
\citep{Kasen2017}. Thus, we only allow the energy to vary by a factor of
three, assuming that all of the ejected matter shown in the kilonova
emission has the same energy distribution. For the $n_{_{\rm ISM}}$ we
allow a range of $-2.4\, >\,\log_{10}(n_{_{\rm ISM}})\,>\,-4.9 $, which
comes from 3-$\sigma$ uncertainties (see Fig. \ref{fig:corner}). If
future observations show a deviation from the jet afterglow (the red
solid line in Fig. \ref{fig:eject-fut}) at very late times, this would
signal the non-thermal appearance of the ejected matter, that was
responsible for the thermal kilonova emission.

\section{Conclusions}
\label{sec:con}

We have performed 3D general-relativistic MHD simulations to model the
self-consistent launch of a jet after a BNS merger. Our initial
configuration consists of a rotating black hole surrounded by a matter
distribution that is inspired from numerical-relativity simulations where
the merger process is modeled from first principles and that is meant to
represent the matter ejected at the merger and till black-hole
formation. We follow the evolution of the outflow beyond the break-out
from the merger ejecta and use the results to extract the a
self-consistent angular structure of the relativistic outflow. The
outflow structure found in this way from the 3D simulation provides an
important confirmation on the previous results obtained from 2D
simulations, where it was shown that relativistic MHD outflows breaking
out from merger ejecta have a hollow core of $\theta_{\rm
  core}\approx4^{\circ}$ and an opening angle of $\theta_{\rm
  jet}\gtrsim10^{\circ}$ \citep{Nathanail2020b}.

The results of the simulations have also been used to calculate the
expected non-thermal afterglow emission and to fit the panchromatic
afterglow from GRB170817A that was observed between $9-940\,{\rm d}$ post
merger \citep{Makhathini2020}. During the fitting procedure we ensure
that the afterglow emission is consistent with the VLBI constraints for
the superluminal motion \citep{Mooley2018b,Ghirlanda2019}. In this way,
we obtain an observer angle of $\theta_{\rm obs}= 35.7^{\circ
  \,\,+1.8}_{\phantom{\circ \,\,}-2.2}$, consistent with independent
estimations coming from GW and kilonova photometry.

The relativistic outflow in the simulations is accompanied by an MHD wind
of ejected matter that is distributed with velocities of
$0.05\,c\,<\,v\,<\,0.9\,c$, with most of the ejected mass moving at
$v<0.3c$, again, a result consistent with inferences derived from
kilonova observations. This component of the ejected matter is expected
to give rise to a thermal emission and to produce the kilonova signal,
which however we do not model in our study. On the other hand, we use the
energy distribution of the ejected matter to compute the non-thermal
afterglow emission that comes from the ejected matter. In this way, we
conclude that this non-thermal emission from the ejected matter is
potentially detectable from future observations and we provide
constraints on the maximum flux density that could be expected and on the
allowed parameter space for the observations.

\section*{Acknowledgements}
Support comes in part also
from ``PHAROS'', COST Action CA16214 and the LOEWE-Program in HIC for
FAIR. The simulations were performed on the SuperMUC cluster at the LRZ
in Garching, on the LOEWE cluster at the CSC in Frankfurt, and on the
HazelHen cluster at the HLRS in Stuttgart. RG's research is supported by 
the ISF-NSFC joint research program (grant No. 3296/19).

\section*{Data Availability}
The data underlying this article will be shared on reasonable request to
the corresponding author.




\bibliographystyle{mnras}
\bibliography{3D-MHDjet.bbl}



\appendix
\section{On the development of the MRI}
\label{appen}

\begin{figure}
  \centering
  \includegraphics[width=0.75\columnwidth]{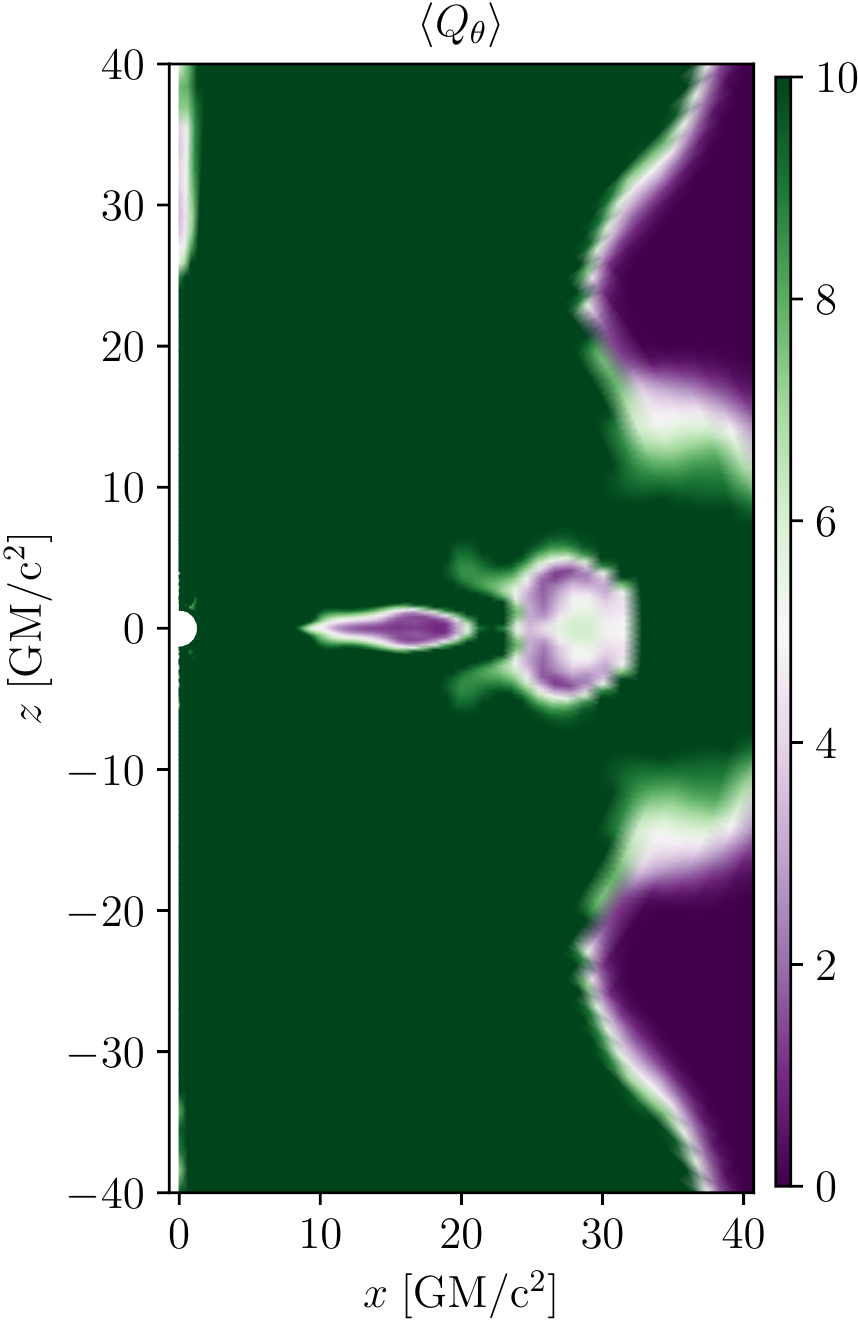}
  \caption{The MRI quality factor $Q_{\theta}$ in the inner
      regions of the accretion disc and close to the black hole.}
  \label{fig:MRI}
\end{figure}

As mentioned in the main text, the initial torus configuration
  leads to the development of the MRI and through the transfer of angular
  momentum drives the accretion on to the torus. Since this is an
  important aspect of the simulations reported here, it is important to
  validate that the resolution employed is sufficient to resolve properly
  the MRI and hence ensure that the accretion process is realistic.  ,
  thus there is the need to properly resolve the MRI inside the torus. As
  customary in these cases, the validation is realised in terms of the
  $\phi$-averaged quality factor for the MRI,
  $Q_{\theta}:=\lambda_{\theta}/dx_{\theta}$, where $\lambda_{\theta}$ is
  the wavelength of the fastest growing mode for the MRI and
  $dx_{\theta}$ the corresponding grid spacing \citep{Porth2019}. This
  quantity is reported in Fig. \ref{fig:MRI} as measured in the inner
  parts of the accretion disc and hence in the vicinity of the black
  hole. As indicated by the colorcode, $Q_{\theta} \gtrsim 10$, hence
  indicating that the MRI is properly resolved in the relevant region.

%

\end{document}